# Unusual nanoscale coexistence of polar–nonpolar domains underlying oxygen storage properties in Ho(Mn, Ti)O$_{3+\delta}$

Hiroshi Nakajima[1], Kento Uchihashi[1], Hirofumi Tsukasaki[1], Daisuke Morikawa[2], Hiroyuki Tanaka[3], Tomohiro Furukawa[3], Kosuke Kurushima[4], Jun Yamasaki[5], Hiroki Ishibashi[6], Yoshiki Kubota[6], Atsushi Sakuda[3], Akitoshi Hayashi[3], and Shigeo Mori[1]

[1]*Department of Materials Science, Osaka Metropolitan University, Sakai, Osaka 599-8531, Japan*

[2]*Institute of Multidisciplinary Research for Advanced Materials, Tohoku University, Sendai 980-8577, Japan*

[3]*Department of Applied Chemistry, Osaka Metropolitan University, Sakai, Osaka 599-8531, Japan*

[4]*Toray Research Center, Ohtsu, Shiga 520-8567, Japan*

[5]*Research Center for Ultra-High Voltage Electron Microscopy, Osaka University, Ibaraki 567-0047, Japan*

[6]*Department of Physics, Graduate School of Science, Osaka Metropolitan University, Sakai, Osaka, 599-8531, Japan*

ABSTRACT

Hexagonal manganese oxides $R$MnO$_3$ show intriguing topological ferroelectric-domain walls with variable conductivity, leading to domain wall engineering. Despite the numerous experimental studies on the polar nanoscale structures, controlling ferroelectric domains has not been sufficiently investigated. Here, we reveal the unprecedented coexistence of polar–nonpolar nanoscale domains that can be formed by substituting Ti ions in HoMnO$_3$. Unusual polar nanoscale domains are embedded in nonpolar domains with different crystallographic symmetry. This polar–nonpolar coexisting structure is naturally assembled by adjusting the lattice length during a solid-state reaction process. Furthermore, the comprehensive study reveals that the reversible microstructural change with a nonpolar–polar transition is strongly correlated with the oxygen storage properties in Ho(Mn, Ti)O$_{3+\delta}$. The present results provide important insight into the nanoscale polar–nonpolar domain coexistence in functional rare-earth manganese oxides, $R$MnO$_3$.

Ferroelectric materials generate reversible spontaneous polarization, which makes the development of numerous electric devices possible, such as storage media and sensors. Particularly, ferroelectric boundaries are useful for fabricating devices because of their small size (nanoscale size) and polarization switching by electric fields [1,2]. Furthermore, some ferroelectric-domain boundaries have characteristic conductivities that are different from those of the bulk [3]. However, forming nanosized polar domains in ferroelectrics is difficult because the electric field should be locally applied by using a scanning probe. Although some ferroelastic oxides exhibit a polar nature at their twin boundaries [4–6], the size of the polar boundaries cannot be adjusted in oxide synthesis processes. Therefore, materials that naturally form ferroelectric nanodomains are desirable for nanoelectronics. Further, it is preferable that the polar domain and boundary size can be controlled by varying the oxide synthesis conditions. To explore a new type of nanoscale polar configurations, we focus on the ferroelectricity of $R$MnO$_3$ ($R$ = Y, Ho–Lu, Sc, and In), which exhibits intriguing ferroelectric structures because of Mn trimerization and ionic displacement of rare-earth elements [7,8].

In the bulk, $R$MnO$_3$ crystallizes in a hexagonal crystal structure of the space group $P6_3cm$ at room temperature and $P6_3/mmc$ at the high-temperature phase. Hexagonal $R$MnO$_3$ causes various intriguing physical phenomena. It exhibits multiferroic properties of antiferromagnetism and ferroelectricity, and these two types of domains are correlated within a crystal [9,10]. Moreover, the material system exhibits peculiar cloverleaf ferroelectric domains accompanying head-to-head or tail-to-tail polarized domain walls and antiphase boundaries [11,12]. The charged domain walls present in cloverleaf domains offer a unique opportunity to manipulate conductivity by accumulating holes [13,14], which can be applied in the fabrication of electrical devices using $R$MnO$_3$ [15].

Hexagonal $R$MnO$_3$ is also a promising candidate for practical oxygen storage materials [16]. Particularly, HoMnO$_3$ exhibits excellent reversible incorporation of a large amount of excess oxygen [17,18]. The oxygen incorporation and release processes occur within a narrow and low operating temperature range of 180 °C – 320 °C. The structural analysis indicated that $R$MnO$_{3+\delta}$ with excess oxygen $\delta$ has a rhombohedral structure with a tripling of the lattice constant of the $c$ axis. A recent structural analysis suggests that HoMnO$_3$ has a rhombohedral structure $R\bar{3}c$ due to Ti substitution [19]. Therefore, investigating the microstructure and oxygen storage property of Ti-substituted HoMnO$_3$ is crucial for understanding the functions of the hexagonal system. Although the substitution effects of HoMnO$_3$ have not been as extensively studied as those of pristine $R$MnO$_3$, the introduction of cations with varying valences is expected to induce notable alterations in both the structural and electronic properties of the material. Some previous studies reported that YMnO$_3$ can produce characteristic microstructures by Ti substitution; however, detailed analyses are lacking [20–22].

This study reveals the nanoscale structural changes in the polar domains of the Ti-substituted HoMnO$_3$. The local observation revealed an unexplored polar boundary structure, whose size could be changed by manipulating the Ti content. This study proposes a strategy for fabricating aligned polar boundaries within a nonpolar matrix. Furthermore, the study demonstrates that the composite structure can be changed dynamically during the oxygen release process. The results are critical for understanding the nanoscale structures that determine various physical and chemical properties of functional ferroelectric materials.

## 2. Experimental methods

HoMn$_{1-x}$Ti$_x$O$_{3+\delta}$ ($0 \leq x \leq 1.0$) polycrystalline samples were prepared by a solid-state reaction. Powders of Ho$_2$O$_3$, Mn$_2$O$_3$, and TiO$_2$ were weighed in molar ratios and mixed with ethanol. The specimens were then formed into pellets by using a hydraulic press and baked at 1400 °C for 24 h under atmospheric conditions. Subsequently, the specimens were dry-mixed in a mortar, formed into pellets again, fired at 1400 °C for 24 h, and cooled to room temperature in air. Powder x-ray diffraction (XRD) was performed with a SmartLab (Rigaku Co., Ltd.) using the Cu $K_\alpha$ wavelength. The amount of each phase was evaluated by fitting the XRD profiles through the JANA2006 software [23]. The synthesized specimens were characterized by energy-dispersive x-ray spectroscopy (EDS). The EDS maps showed that the Ti ions were homogeneously located at Mn sites (see Supplementary Fig. 1) [24].

Selected-area electron diffraction, dark-field imaging, and high-resolution transmission electron microscopy (HR-TEM) observations were made using JEM-2010 and JEM-2100F (JEOL Co., Ltd.) at an acceleration voltage of 200 kV. High-angle annular dark-field scanning TEM (HAADF-STEM) was performed using the JEM-ARM200F with a probe semiangle of 22 mrad and a current of 60 pA. EDS was performed with a solid angle of approximately 1.96 sr to analyze elemental mapping [25]. The temperature was raised from room temperature to 440 °C using a heating TEM holder. Polycrystalline specimens were crushed and dispersed onto a carbon grid with ethanol. Electron diffraction patterns were analyzed through a simulation program [26].

Thermogravimetry (TG) measurements were performed with 74.3 mg of the powder specimens on a platinum pan with flowing nitrogen (N$_2$) gas or air (Rigaku, Thermo Plus EVO2 TG-DTA8122) at a rate of temperature increase of 10 °C/min.

## 3. Results and discussion

First, we revealed the phase change due to Ti substitution. Figure 1(a) shows the powder XRD profiles of HoMn$_{1-x}$Ti$_x$O$_{3+\delta}$ ($0 \leq x \leq 1.0$). The material underwent a systematic structural change after Ti substitution. Bragg reflections of $x \leq 0.1$ can be indexed by the hexagonal $P6_3cm$ space group. On increasing $x$, the specimens for $0.2 \leq x \leq 0.4$ demonstrated different peak positions, which can be indexed with the $R\bar{3}c$ structure [19]. Further increase in Ti content resulted in the formation of $R\bar{3}c$ and $Fd\bar{3}m$ structures at $0.5 \leq x \leq 0.8$, indicating that the two phases coexisted in the average structure. At $x \geq 0.9$, the XRD patterns possessed a single phase $Fd\bar{3}m$ structure, which is the pyrochlore structure of Ho$_2$Ti$_2$O$_7$. The Rietveld analysis based on the two phases revealed that the amount of the pyrochlore phase is 0.30 at $x = 0.5$, 0.55 at $x = 0.6$, 0.65 at $x = 0.7$ and 0.8, 1.0 at $x \geq 0.9$, demonstrating that the pyrochlore $Fd\bar{3}m$ phase increases while the $R\bar{3}c$ phase decreases with increasing Ti substitution.

To further reveal the structural changes, selected-area electron diffraction patterns were obtained along the [110] axis for each phase, as shown in Fig. 1(b). At $x = 0$, the specimen showed a single-crystal $P6_3cm$ pattern. In contrast, the electron diffraction at $x = 0.2$ showed a complex pattern. A comparison of this complex pattern with diffraction simulations [Supplementary Fig. 2(a)–(d)] demonstrated that the pattern has a $P6_3cm$ symmetry superimposed on two twin-related $R\bar{3}c$ structures: the 0012 reflection of $R\bar{3}c$ was superimposed on the 004 reflection of $P6_3cm$, and the $3\bar{3}0$ reflection was common in both structures [24]. The selected-area pattern was obtained from a single grain. The result indicates that the two phases locally coexisted in a single grain at $0.2 \leq x \leq 0.4$ though the average structure was the rhombohedral $R\bar{3}c$. The detailed composite structure is discussed in the next paragraph. A similar diffraction pattern was obtained for the specimen at $x = 0.7$, suggesting that the local structure has a mixture of $R\bar{3}c$ and $P6_3cm$ in the grain with the average $R\bar{3}c$ structure. The $Fd\bar{3}m$ structure was mixed as another grain for $0.5 \leq x \leq 0.8$. The specimen at $x = 1.0$ was a single phase consistent with the $Fd\bar{3}m$ structure.

The composite structure was examined through real-space dark-field imaging. In Fig. 2, the dark-field images show the characteristic meandering boundaries of the $P6_3cm$ regions. Since these images were captured using the $2\bar{2}0$ reflection of $P6_3cm$, bright boundaries correspond to the $P6_3cm$ structure, and dark domains have the $R\bar{3}c$ structure. The domain size of $R\bar{3}c$ increased, and the area density of bright $P6_3cm$ lines decreased from $x = 0.2$ to 0.3. Conversely, the $R\bar{3}c$ domains were again refined as the $x$ further increased from 0.4 to 0.7, suggesting that the $R\bar{3}c$ structure may be energetically the most stable at around $0.3 < x < 0.4$. This stability is inferred from Fig. 1(a) showing that the pyrochlore phase of the $Fd\bar{3}m$ structure increases above $0.5 \leq x$, while the $x \leq 0.2$ side of the phase diagram has the $P6_3cm$ structure. Besides, the bright $P6_3cm$ regions widen with increasing Ti content, and numerous large regions of $P6_3cm$ are formed, as shown in the dark-field image for $x = 0.7$. Importantly, a composite microstructure of the $P6_3cm$ and $R\bar{3}c$ structures was formed within a single grain in all the compositions: The $P6_3cm$ boundaries are not formed as grain boundaries between different grains.

To reveal the local structures, we performed HR-TEM (Fig. 3) and HAADF-STEM (Fig. 4). Figure 3 shows the observation results of the composite structure at $x = 0.3$. This area also shows the two-phase mixture, as illustrated in the dark-field image of Fig. 3(a). The HR-TEM image of the boundary marked by the yellow arrow was obtained. The HR-TEM image given in Fig. 3(b) reveals that the boundary is arranged without a void region. The fast Fourier transform (FFT) patterns of areas A, B, and C are depicted in the right panels. The dotted rectangles in the lattice image and FFT patterns show that the structures have a mirror-symmetrical relationship, demonstrating that the $R\bar{3}c$ domains have twin domains. Because of the twin formation, bright areas were changed and the contrast was reversed when the dark-field images were captured using different orientational spots, as shown in Fig. 3(c). The contrast reversal and twin domains of $R\bar{3}c$ with the $P6_3cm$ boundaries were also discovered in a wide range of $0.15 \leq x \leq 0.7$ (See Supplementary Figs. 3–7) [24]. The FFT pattern of area B shows a single phase of $P6_3cm$ at the boundary. Furthermore, the orientation and lattice constants of the $R\bar{3}c$ structure matched those of $P6_3cm$. This match in the lattice constants and orientation was also confirmed by the electron diffraction patterns without any peak splitting, as shown in Figs. 1 and 2.

The highly aligned boundary was realized by changing the lattice constants. The bulk specimen has the following lattice constants: $a = 6.141$ Å and $c = 11.42$ Å for $P6_3cm$, and $a = 6.241$ Å and $c = 33.38$ Å for $R\bar{3}c$ [19]. Note that the comparable lattice constant is $c = 11.13$ Å for $R\bar{3}c$ since the lattice constant of the $c$ axis is threefold in the structure. The lattice constants ($a$ and $c$ axes) of $P6_3cm$ are shorter and longer than those of $R\bar{3}c$, respectively. The electron diffraction patterns based on the lattice constants of the bulk crystals produced a split zigzag diffraction pattern, which did not agree with the experimental diffraction pattern (See Supplementary Fig. 2(e)) [24]. When the lattice constants of the $a$ and $c$ axes in $P6_3cm$ were changed to those of $R\bar{3}c$, the simulation pattern reproduced the experimental diffraction pattern, as shown in Supplementary Fig. 2(f) [24]. Therefore, at the $P6_3cm$ boundary, the lattice constant of the $a$ axis stretches and that of the $c$ axis shrinks to form an atomically aligned boundary.

The matching of the lattices in the two structures was also maintained at wavy boundaries, as revealed in the HAADF-STEM image shown in Fig. 4(a). Although the crystal structure of $P6_3cm$ (Fig. 4b) was different from that of $R\bar{3}c$ (Fig. 4c), the boundaries were smooth and atomically sharp, as indicated by the yellow lines. Noticeably, the $P6_3cm$ area maintained the up-down-down displacements of Ho atoms similar to those in bulk $R$MnO$_3$ [27], demonstrating that the boundary has spontaneous ferroelectric polarization. On the contrary, the $R\bar{3}c$ domains had the up-stay-down shifts of Ho atoms without polarization. The convergent-beam electron diffraction patterns shown in Supplementary Fig. 8 further support the nonpolar space group [24, 28]. Because the $R\bar{3}c$ domains have a centrosymmetric structure without polarization and the $P6_3cm$ boundaries exhibit ferroelectric polarization due to the shift of Ho atoms, this composite structure exhibits ferroelectric polarization only at thin boundaries. Recently, two-dimensional polar structures have garnered much attention in terms of domain wall engineering of memory devices [1]. HoMnO$_3$ exhibits large spontaneous polarization (approximately 5 µC/cm$^2$) [29,30]. Besides, the polar $P6_3cm$ boundaries are formed naturally by the conventional solid-state reaction. Hence, these results demonstrate that the Ti-substituted HoMnO$_3$ is beneficial for fabricating ferroelectric nanodomains.

Comparing the polar boundary with other polar ones is useful for highlighting the importance of the structure. In CaTiO$_3$, polar boundaries exist between twin domains [4,5]. Although the bulk CaTiO$_3$ has a nonpolar space group of $Pnma$, the twin boundary has point-group symmetry of $m$ or 2, resulting in the polar nature observed at the boundary. These polar boundaries have specific crystallographic orientations because of the twin relationships. Besides, the boundaries are straight because of the presence of a crystal plane. Conversely, the polar boundaries observed in Ho(Mn, Ti)O$_{3+\delta}$ have meandering boundary planes, as can be seen from Fig. 4(a). The curved shape of the polar boundaries indicates that there is no restriction in creating an interface between the different domains of $P6_3cm$ and $R\bar{3}c$. This interface formation in any direction results from the similarity of their ionic positions for Ho, Mn, and O atoms when the crystallographic axes and lattice constants agree with each other. Furthermore, the ionic displacements of $CaTiO_3$ are less than 0.1 Å, suggesting that the polarization at the boundary is small [4]. Figure 4 shows that the ionic displacement of Ho is approximately 0.3 Å in Ti-substituted $HoMnO_3$; this value is comparable to that of pure $HoMnO_3$ [19]. This finding indicates that the spontaneous polarization of the boundary is equivalent to that of the bulk $HoMnO_3$ (5 $\mu C/cm^2$). Consequently, these results prove that the observed polar boundary of $Ho(Mn, Ti)O_{3+\delta}$ has unique characteristics as a polar boundary between nonpolar matrixes.

Furthermore, because pure $HoMnO_3$ in a rhombohedral structure exhibits oxygen absorption and release characteristics through a temperature swing, we investigated the oxygen storage properties and related structural change in $HoMn_{0.7}Ti_{0.3}O_{3+\delta}$. Figure 5(a) shows the TG result obtained under the $N_2$ condition. In the thermal analysis, the oxygen content decreased at approximately 430 °C from $\delta \approx 0.15$ to 0.10, suggesting the release of oxygen from the oxide. When the temperature was decreased from 800 °C to room temperature, no peak was observed because of the $N_2$ atmosphere. The specimen used for the TG measurement under the $N_2$ condition was measured under the air flowing [Fig. 5(b)]. The specimen showed oxygen absorption from room temperature to 450 °C and then the oxygen content decreased to achieve $\delta \approx 0.115$ at 800 °C. Contrary to the $N_2$ condition, the oxygen content increased with the decreasing temperature under the air, demonstrating the oxygen absorption process under the air atmosphere. The oxygen content $\delta \approx 0.131$ at room temperature after the cycle was close to the as-grown state value of $\delta \approx 0.15$. The results indicate the reversible oxygen absorption and release properties of this material. Furthermore, the gas chromatogram in Fig. 5(c) reveals an oxygen release at 430 °C. This temperature is higher than that of oxygen-annealed $HoMnO_3$ by 100 °C [17]. Note that the previous studies used pure $HoMnO_3$ annealed in oxygen; Ti-substituted $HoMnO_3$ was not investigated. Therefore, this study shows that Ti-substituted $HoMnO_3$ prepared in air is a promising oxygen storage material similar to nonsubstituted materials annealed in oxygen.

To correlate the oxygen release process with the microstructure, we performed in situ heating observations. Figure 6 shows the morphological changes during the heating process. At room temperature, the specimens with $x = 0.3$ and 0.6 showed the wavy composite structures of $R\bar{3}c$ and $P6_3cm$, which are similar to the aforementioned observation results. Additionally, the diffraction patterns revealed reflections from both symmetries. On heating to temperatures greater than 440 °C, the $R\bar{3}c$ reflections disappeared and only the $P6_3cm$ reflections remained. Accordingly, the $P6_3cm$ domains were rearranged with a fine structure, as shown in the dark-field images. The disappearance of the $R\bar{3}c$ domains agrees with the oxygen release properties shown in Fig. 5 because the $R\bar{3}c$ structure can stably exist with excess oxygen. The size of bright areas showing ferroelectric domains at $x = 0.6$ was smaller than that at $x = 0.3$ since the long-range order of the domains got disturbed by increasing Ti content. Noticeably, the fine domains of $P6_3cm$ are different from the cloverleaf domains formed in as-grown $HoMnO_3$ [13]. Therefore, this observation revealed that the $R\bar{3}c$ domains change into tiny $P6_3cm$ domains when oxygen is released through the temperature swing process. Furthermore, the $R\bar{3}c$ domains were able to be regained by annealing the specimen at 800 °C in an air atmosphere (see Supplementary Fig. 9) [24]. This phenomenon agrees with the TG results showing that the air annealing recovered the oxygen content in Fig. 5(b). These observations demonstrate that the oxygen release and absorption processes involve the microstructural change of the $R\bar{3}c$ domains.

## 4. Conclusions

We report the nanosized polar domain boundaries in Ti-substituted $HoMnO_3$ observed using several TEM techniques. Real-space observations revealed that the ferroelectric phase of $P6_3cm$ is embedded within the nonpolar domains of the $R\bar{3}c$ structure, which is naturally formed by a solid-state reaction. The two different crystallographic regions can be aligned well without defects by adjusting the lattice constant. The morphology of the polar–nonpolar structure can be changed by manipulating the Ti content, which demonstrates the controllability of the size and density of the polar boundaries. Moreover, the $R\bar{3}c$ structure of $Ho(Mn, Ti)O_{3+\delta}$ can accumulate and release oxygen gas with temperature swing. The oxygen-storage properties are correlated with microstructural changes in the composite domains. The results provide a deep understanding of the substitution effects in the functional material $HoMnO_3$, which exhibits local conductivity at domain walls and magnetoelectric effects [13,31–33]. Our findings further demonstrate the versatility of $HoMnO_3$ regarding oxygen storage properties and the aligned polar structure that is beneficial for its application in ferroelectric nanoelectronics.

*Ti-Doped Hexagonal YMnO$_3$*, Phys. Rev. B **71**, 14114 (2005).

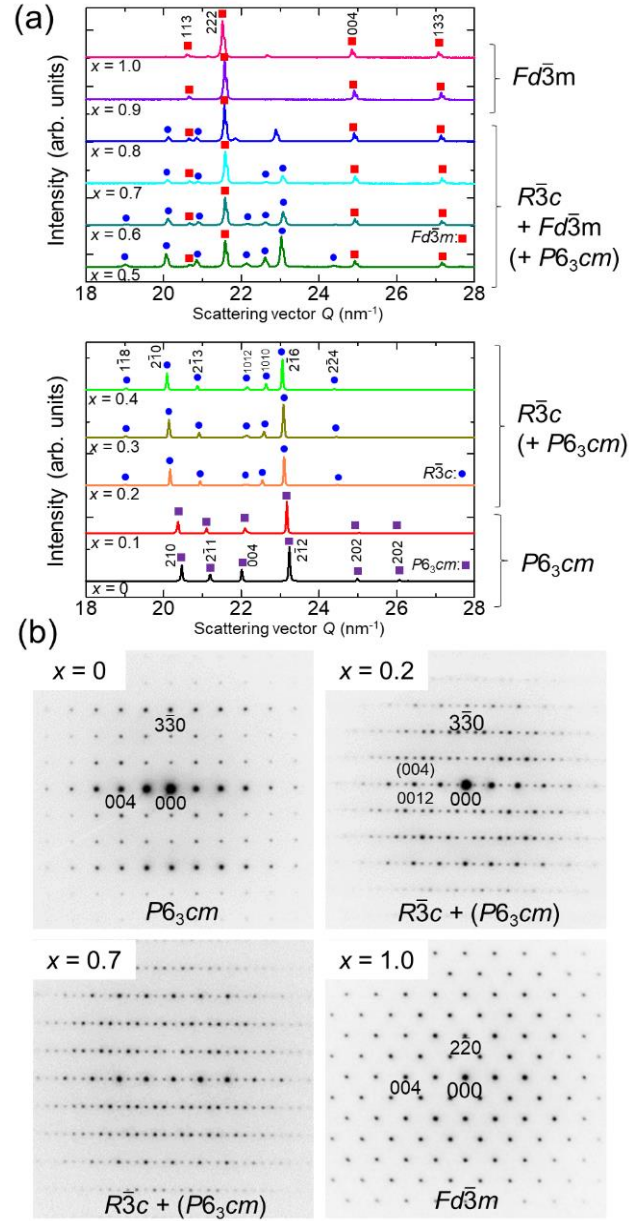

Figure 1. (a) X-ray diffraction (XRD) profile of HoMn$_{1-x}$Ti$_x$O$_{3+\delta}$. The phases of the space groups are listed. The parentheses indicate the local phase that was revealed by transmission electron microscopy (TEM) though the reflection spots are indistinguishable in the XRD profile. (b) Electron diffraction patterns for each phase along the [110] axis. The 0012 and 004 reflections are based on the $R\bar{3}c$ and $P6_3cm$ space groups, respectively. These spots are superimposed on each other.

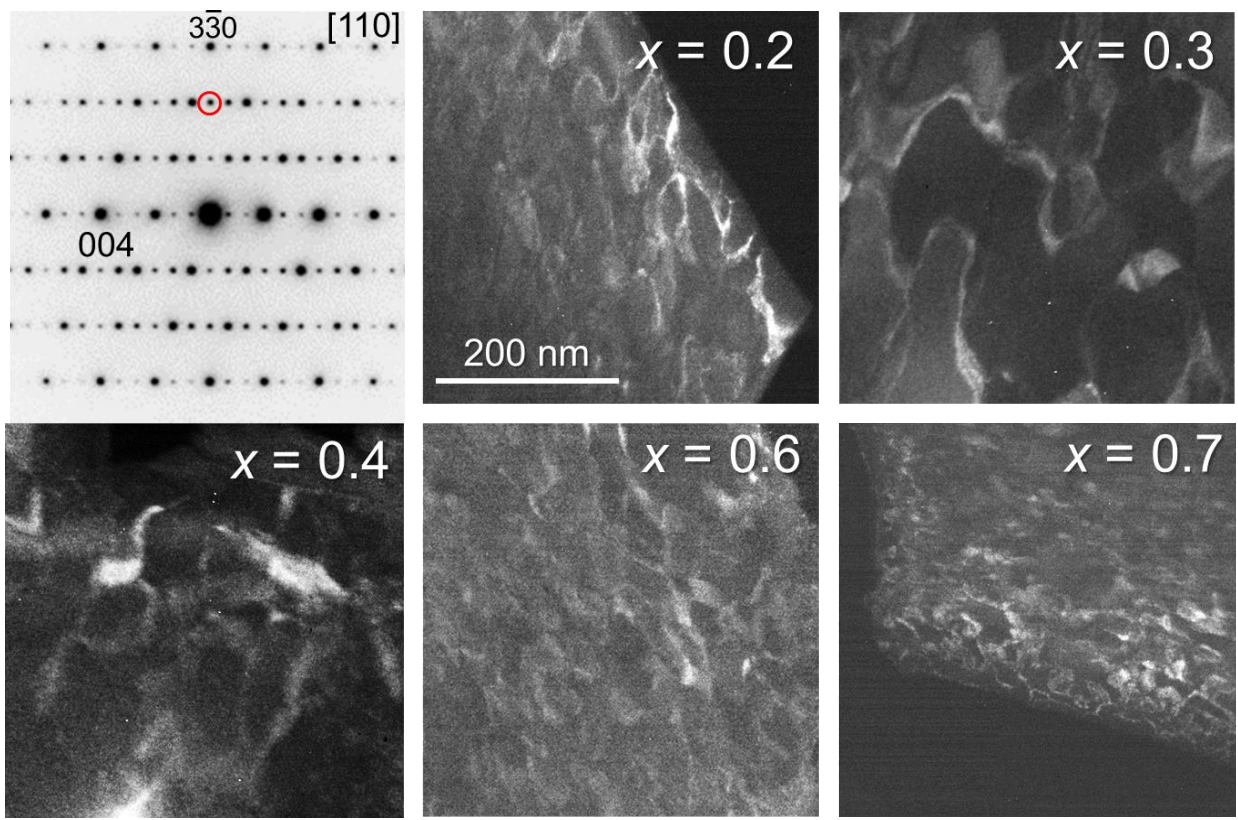

Figure 2. Electron diffraction pattern of the sample with $x = 0.3$ and the dark-field images of the samples with $x = 0.2$–0.7. The dark-field images were obtained from the $2\bar{2}0$ reflection of $P6_3cm$, which is indicated by the red circle in the diffraction pattern. The dark areas depict the $R\bar{3}c$ domains, and the bright lines indicate the $P6_3cm$ regions. All the observations are confined to single grain areas.

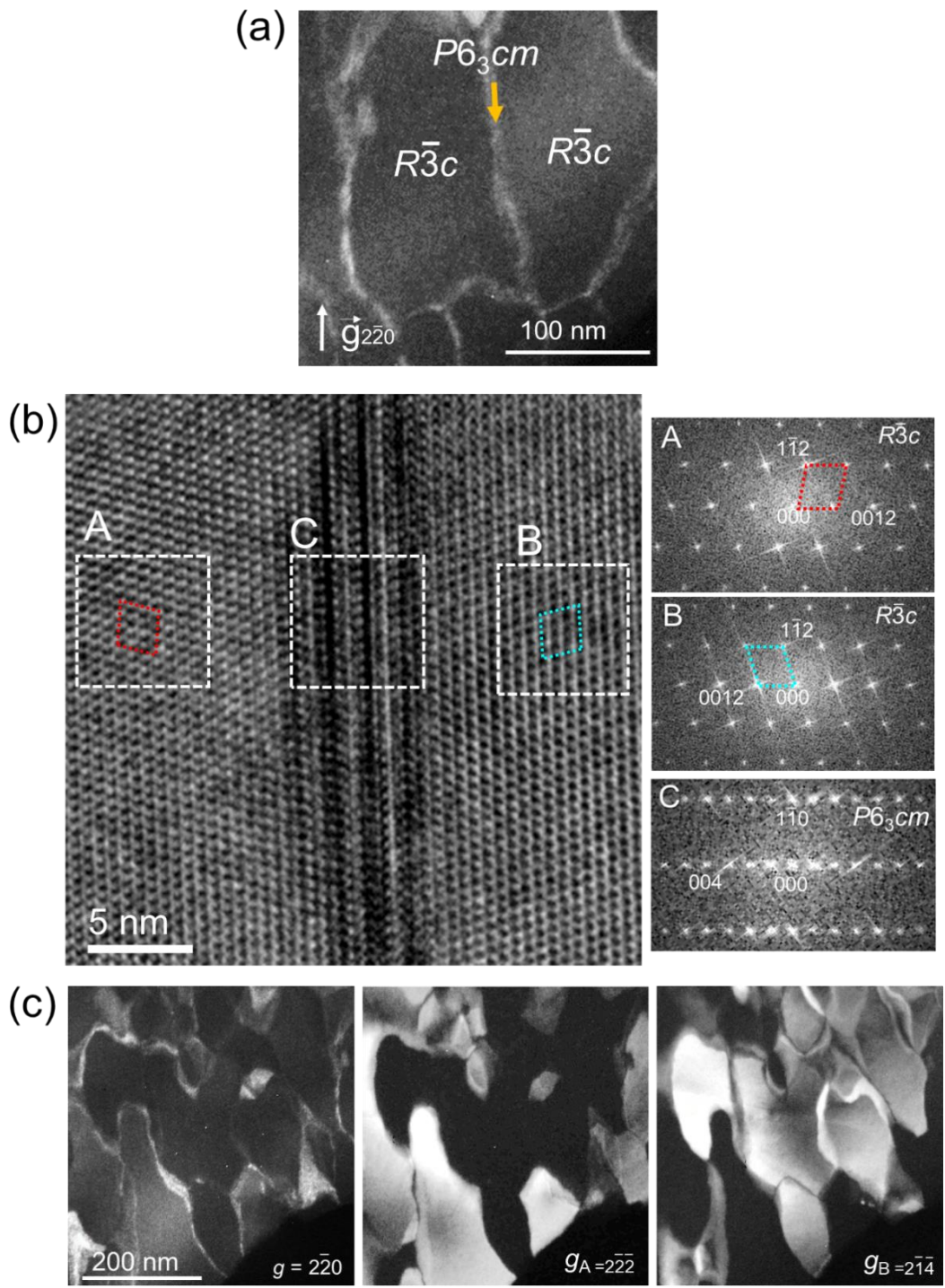

Figure 3. Analysis of the composite structure. (a) Dark-field image of $g = 2\bar{2}0$ reflection (b) High-resolution TEM image of the area indicated by the yellow arrow. The fast Fourier transform (FFT) patterns were calculated from A, B, and C areas. (c) Dark-field images using different Bragg reflections. The contrast is reversed between the images for $g_A$ and $g_B$, which correspond to the two types of domains shown in Supplementary Figure 2(b) and (c), respectively.

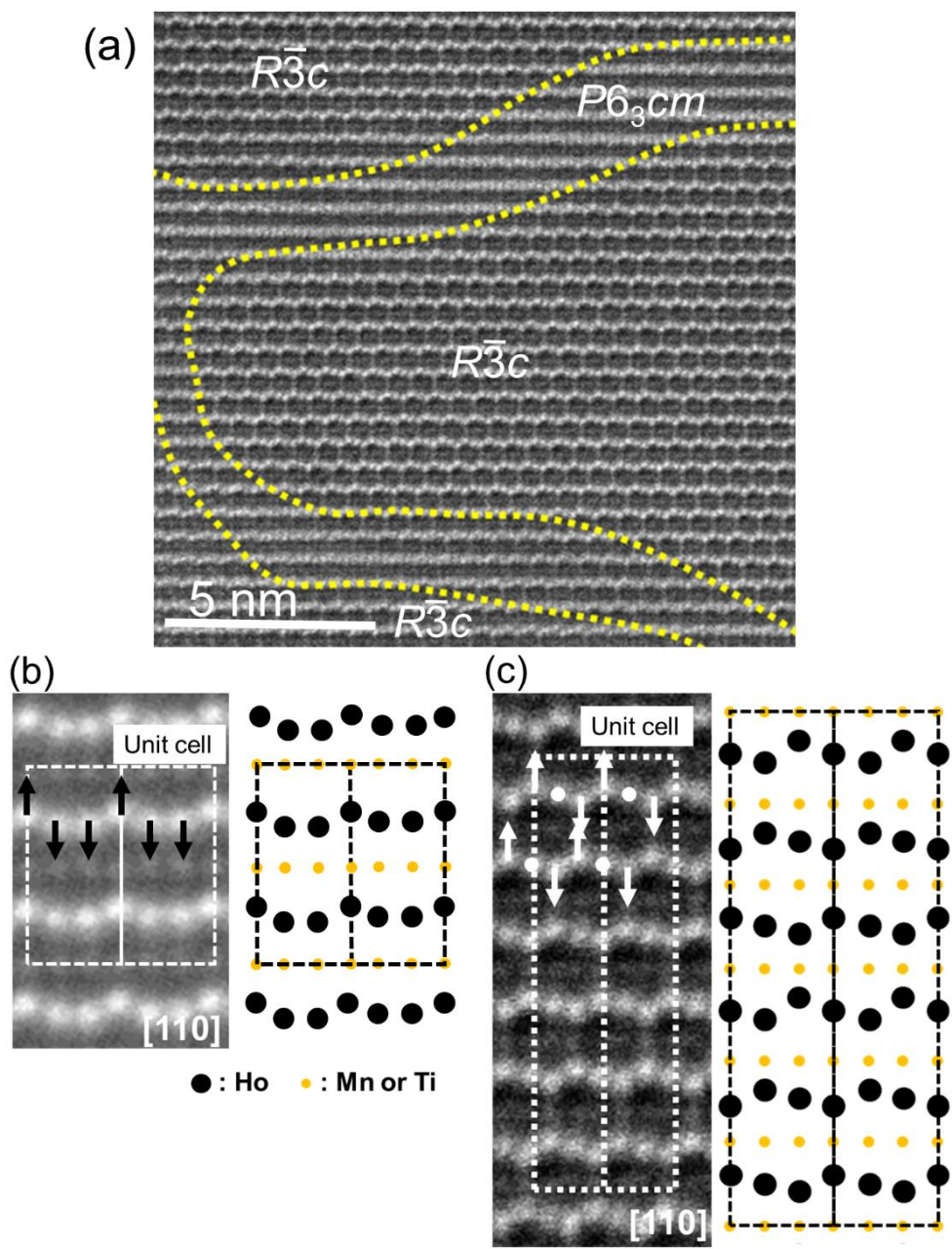

Figure 4. High-angle annular dark-field scanning TEM (HAADF-STEM) image along the [110] axis. The yellow lines show the interface between the $P6_3cm$ and $R\bar{3}c$ areas. (b) Magnified image of a $P6_3cm$ area. The arrows represent the shifts of Ho atoms (the up-down-down structure). (c) Magnified image of an $R\bar{3}c$ area. The arrows indicate shifts of Ho atoms (the up-stay-down structure). The dashed rectangle shows the unit cell in each phase. The schematics show structural models of the HAADF-STEM images.

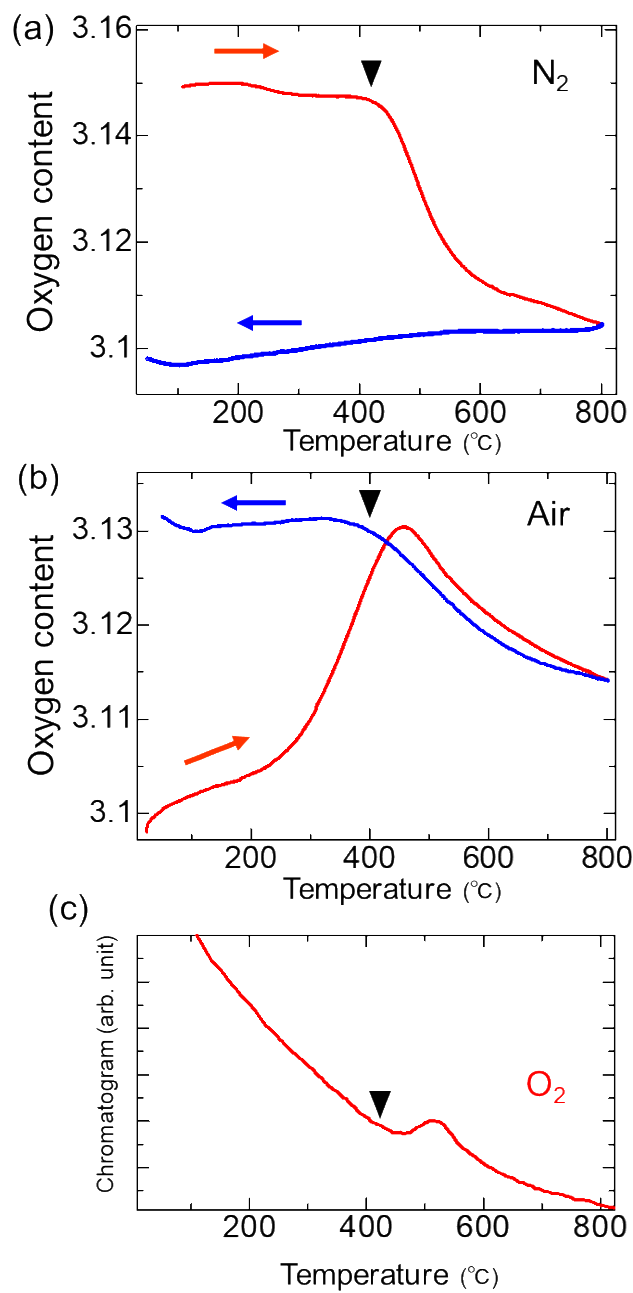

Figure 5. Oxygen storage property of HoMn$_{0.7}$Ti$_{0.3}$O$_{3+\delta}$. (a) Thermogravimetry (TG) measurements under N$_2$. (b) TG measurements under the air. The thermal analysis was conducted in the specimen after the measurement of the N$_2$ atmosphere. (c) Gas chromatography measurement of oxygen on heating. The arrowhead indicates the temperature at which the release of oxygen commenced.

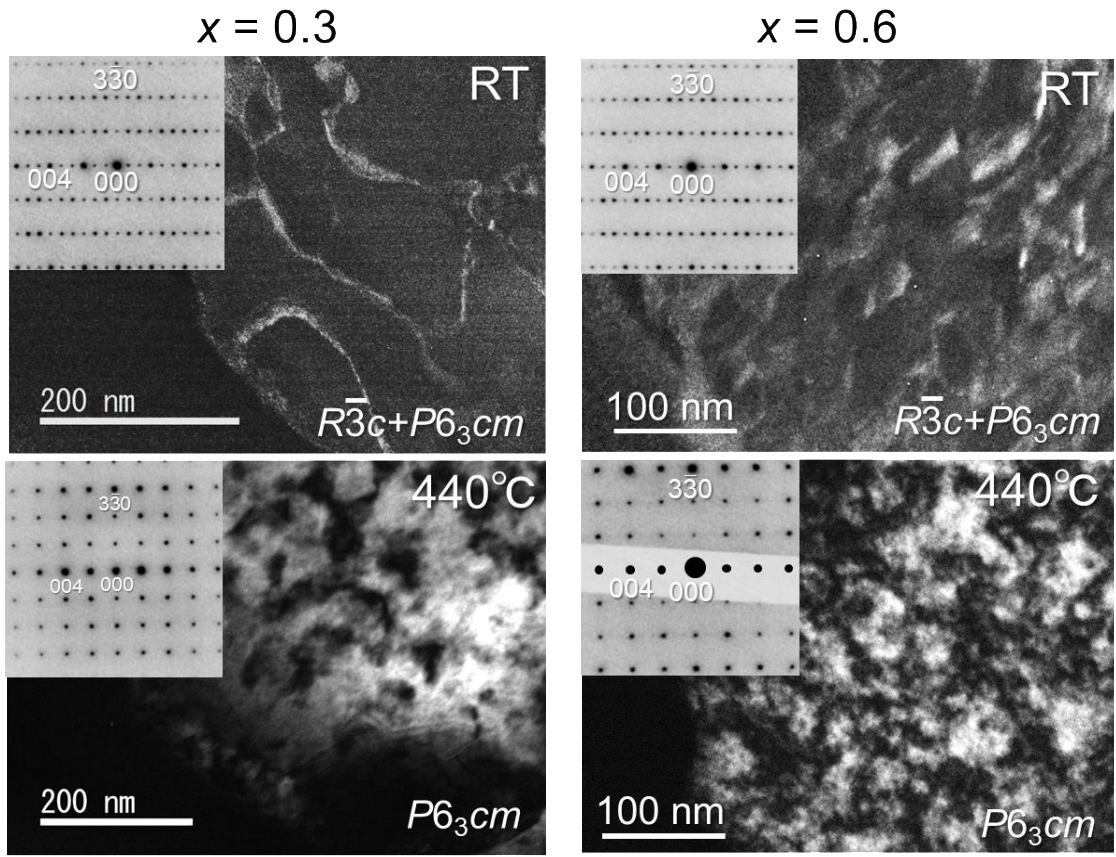

Figure 6. In situ heating observation of HoMn$_{1-x}$Ti$_x$O$_{3+\delta}$ ($x$ = 0.3 and 0.6). The dark-field images for the $2\bar{2}0$ reflection of the $P6_3cm$ space group. The insets in the figures show the electron diffraction patterns of the observed areas.

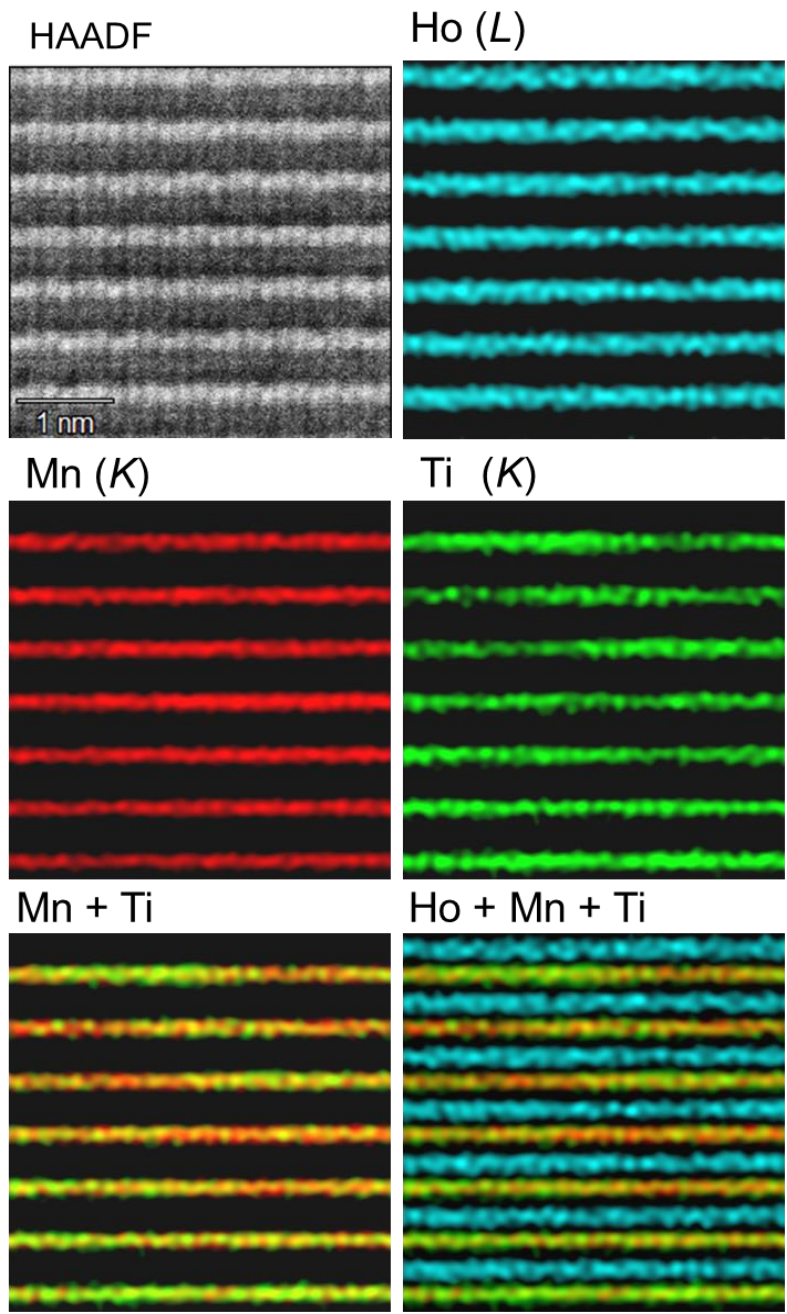

Supplementary Figure 1. HAADF-STEM and energy dispersive x-ray spectroscopy (EDS) maps in $HoMn_{1-x}Ti_xO_{3+\delta}$ ($x = 0.3$). The parentheses indicate the absorption edges used for each elemental mapping. The EDS maps indicate that Ti ions are substituted at Mn sites.

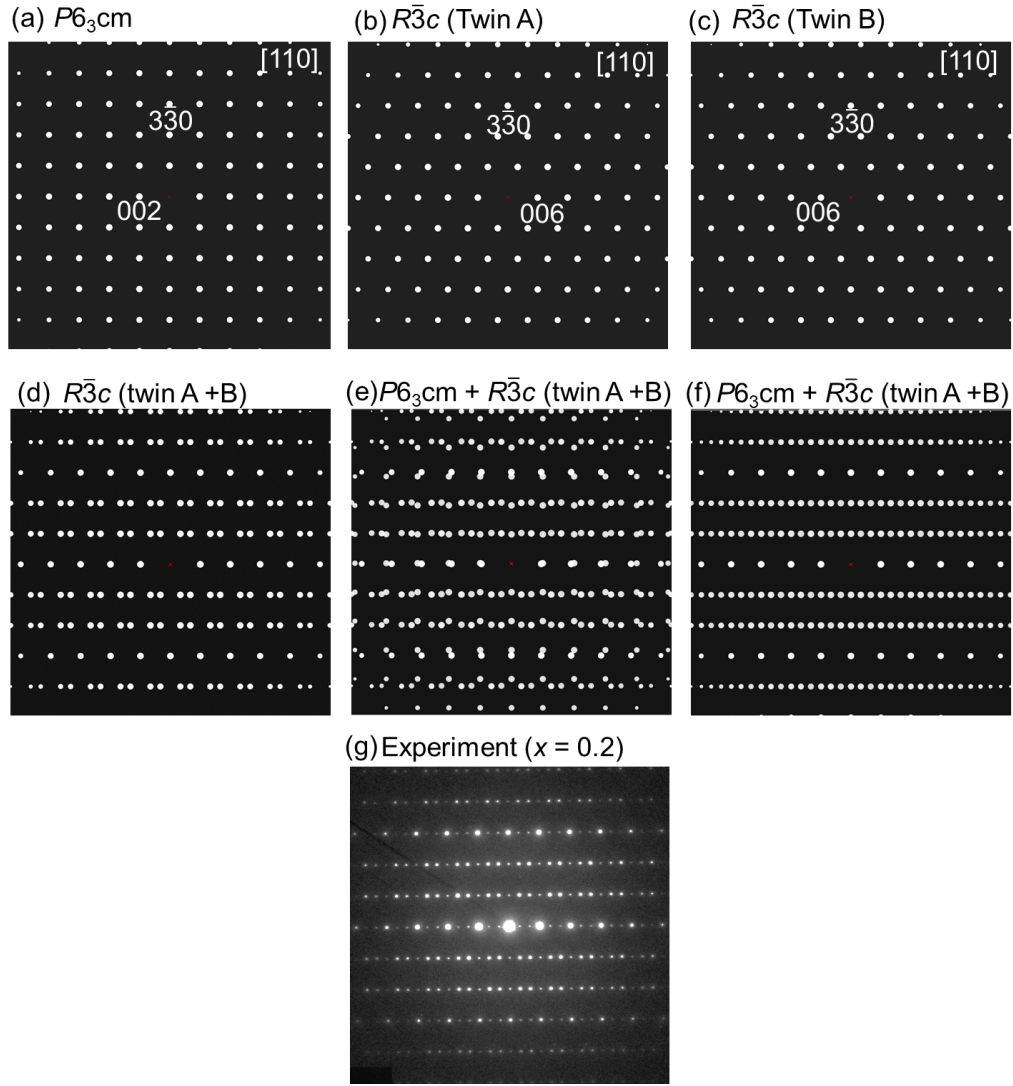

Supplementary Figure 2. (a)–(f) Simulation patterns of electron diffraction along [110]. (a) $P6_3cm$. (b, c) $R\bar{3}c$. The structures of (b) and (c) have a twin relationship. (d) The superimposed pattern of simulated patterns (b) and (c). (e) The superimposed pattern of simulated patterns (a) and (d). The lattice constants of bulks were used for the simulation: $a = 6.141$ Å, $c = 11.42$ Å for $P6_3cm$ and $a = 6.241$ Å, $c = 33.38$ Å for $R\bar{3}c$. (f) The superimposed pattern of simulated patterns (a) and (d) when the lattice constants of the $P6_3cm$ structure agree with those of $R\bar{3}c$. (g) Experimental electron diffraction pattern of $HoMn_{0.8}Ti_{0.2}O_{3+\delta}$, which agrees with the simulation (f). Note that, compared with (f) and (g), the weak reflection spots such as $00L$, and $3\bar{3}L$ ($L$: threefold of the fundamental reflections) appear by double diffraction.

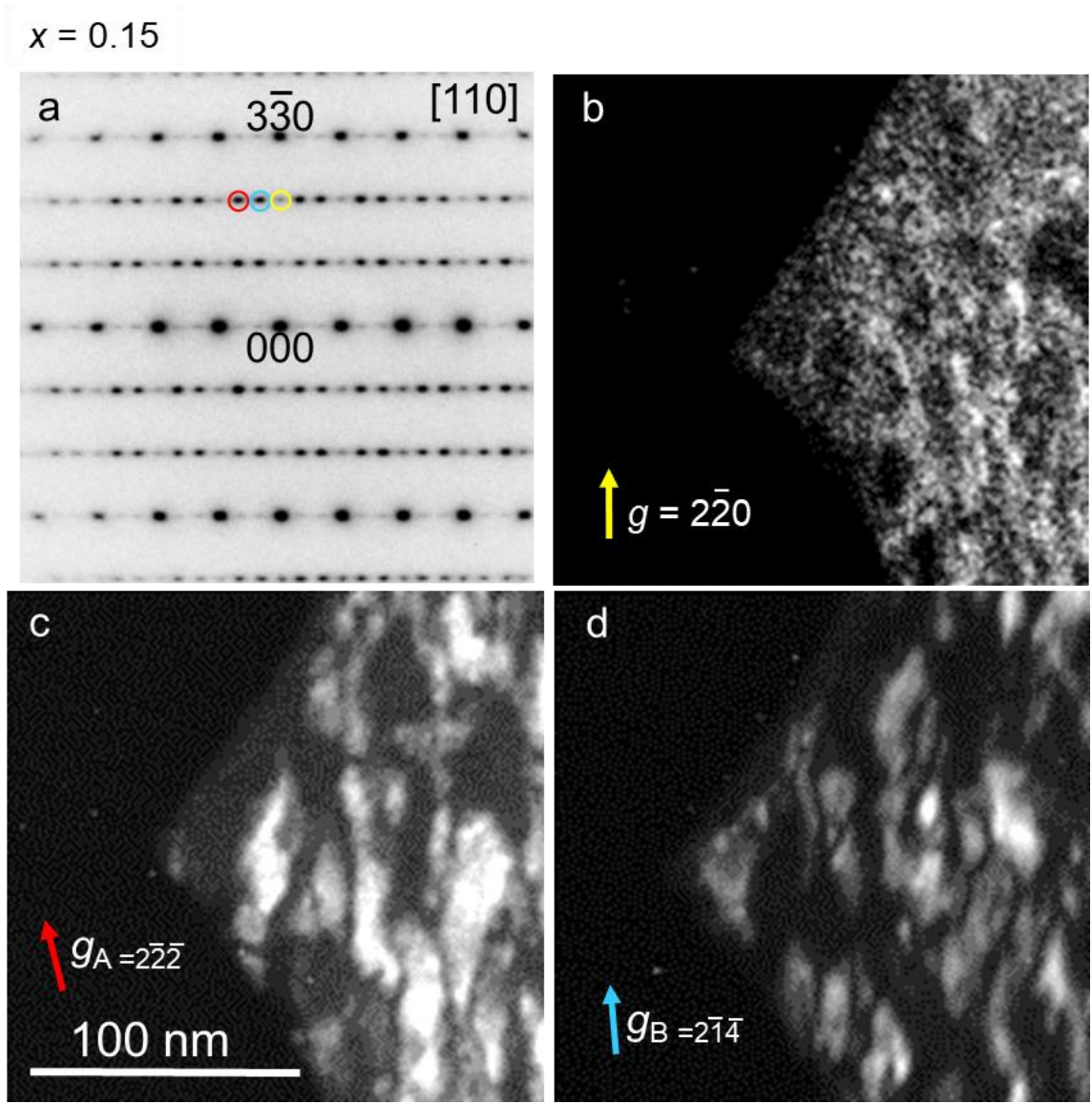

Supplementary Figure 3. Observation of microstructure in HoMn$_{1-x}$Ti$_x$O$_{3+\delta}$ ($x$ = 0.15). (a) Electron diffraction pattern along [110]. The dark-field images were obtained using (b) $g$ = $2\bar{2}0$, (c) $g_A$ = $2\bar{2}\bar{2}$, and (d) $g_B$ = $2\bar{1}\bar{4}$, which are marked by the circles in panel (a). The reflections $g_A$ and $g_B$ originated from the twin domains A and B of Supplementary Fig. 1, respectively.

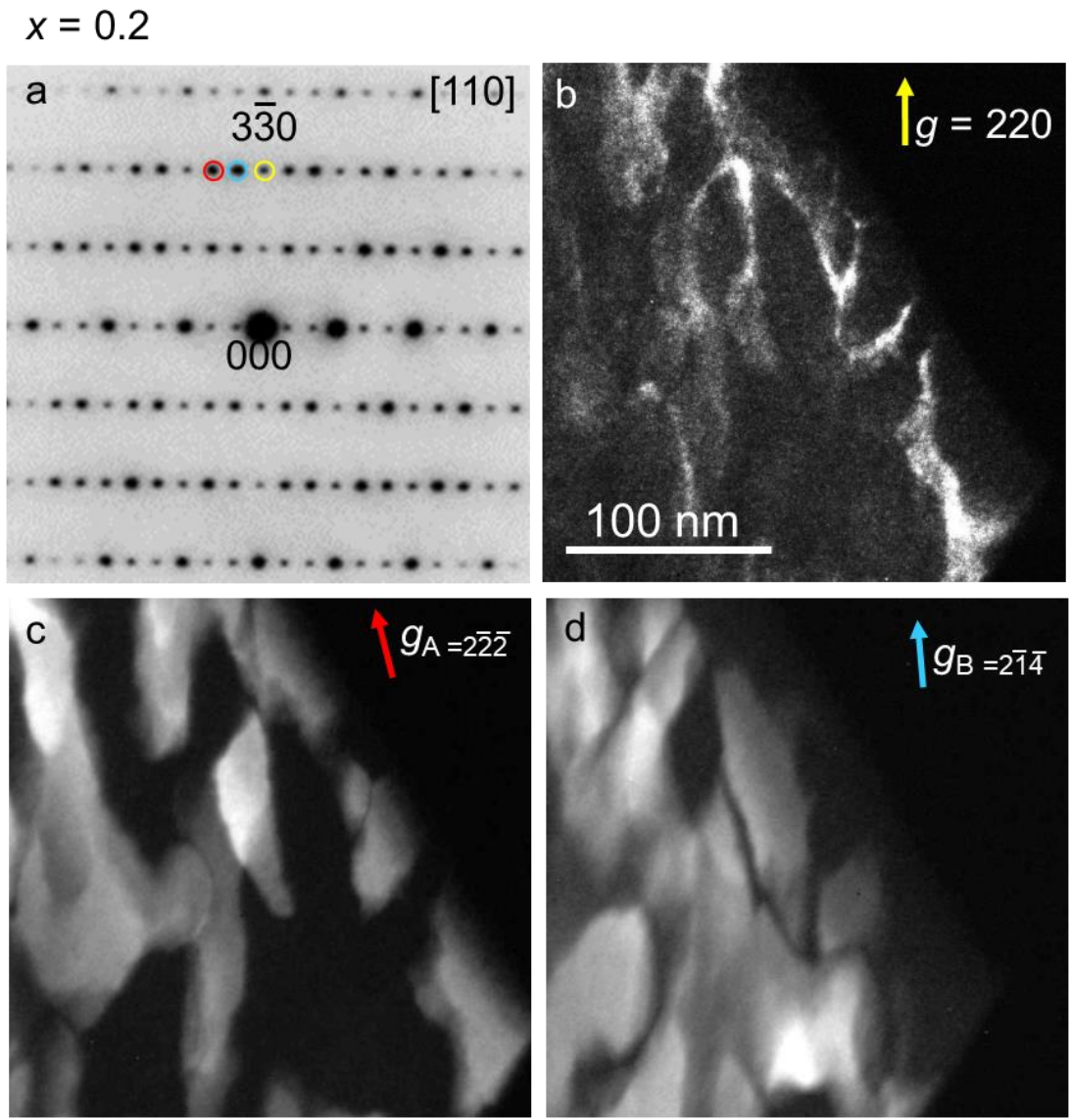

Supplementary Figure 4. Observation of microstructure in $x = 0.2$. (a) Electron diffraction pattern along [110]. The dark-field images were obtained using (b) $g = 2\bar{2}0$, (c) $g_A = 2\bar{2}\bar{2}$, and (d) $g_B = 2\bar{1}\bar{4}$, which are marked by the circles in panel (a). The reflections $g_A$ and $g_B$ originated from the twin domains A and B of Supplementary Fig. 1, respectively.

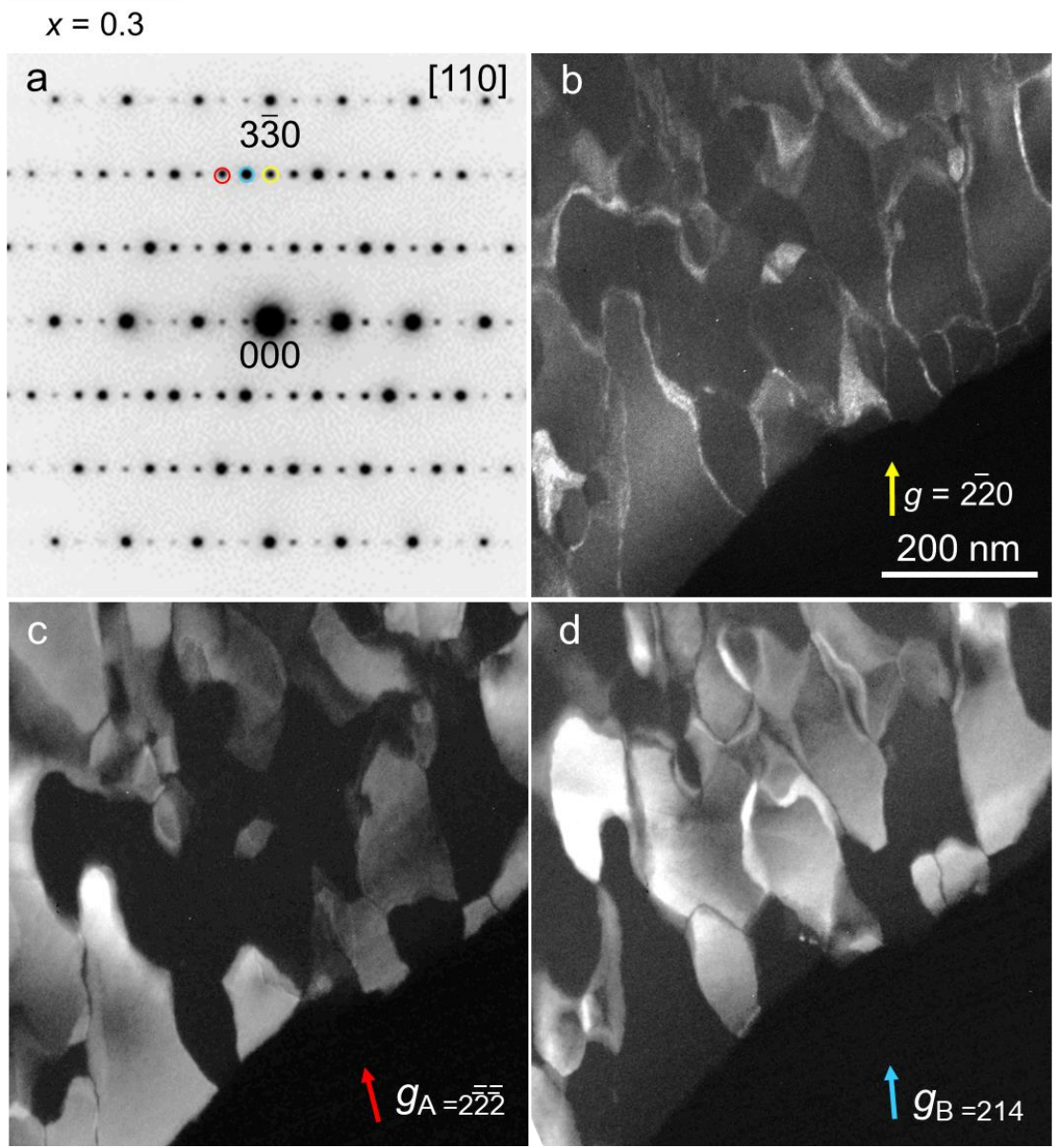

Supplementary Figure 5. Observation of microstructure in $x = 0.3$. (a) Electron diffraction pattern along [110]. The dark-field images were obtained using (b) $g = 2\bar{2}0$, (c) $g_A = 2\bar{2}\bar{2}$, and (d) $g_B = 2\bar{1}\bar{4}$, which are marked by the circles in panel (a). The reflections $g_A$ and $g_B$ originated from the twin domains A and B of Supplementary Fig. 1, respectively.

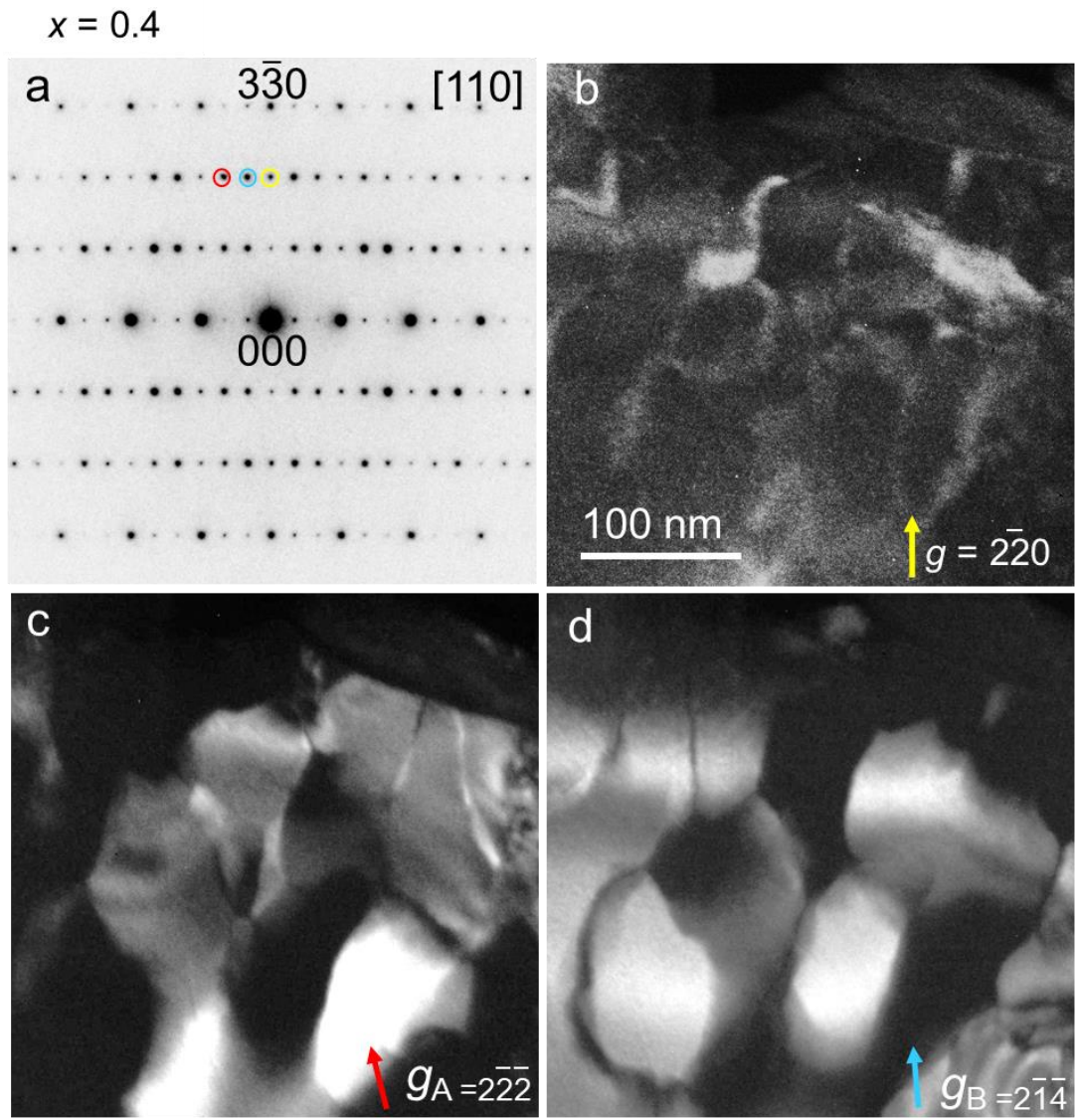

Supplementary Figure 6. Observation of microstructure in $x = 0.4$. (a) Electron diffraction pattern along [110]. The dark-field images were obtained using (b) $g = 2\bar{2}0$, (c) $g_A = 2\bar{2}\bar{2}$, and (d) $g_B = 2\bar{1}\bar{4}$, which are marked by the circles in panel (a). The reflections $g_A$ and $g_B$ originated from the twin domains A and B of Supplementary Fig. 1, respectively.

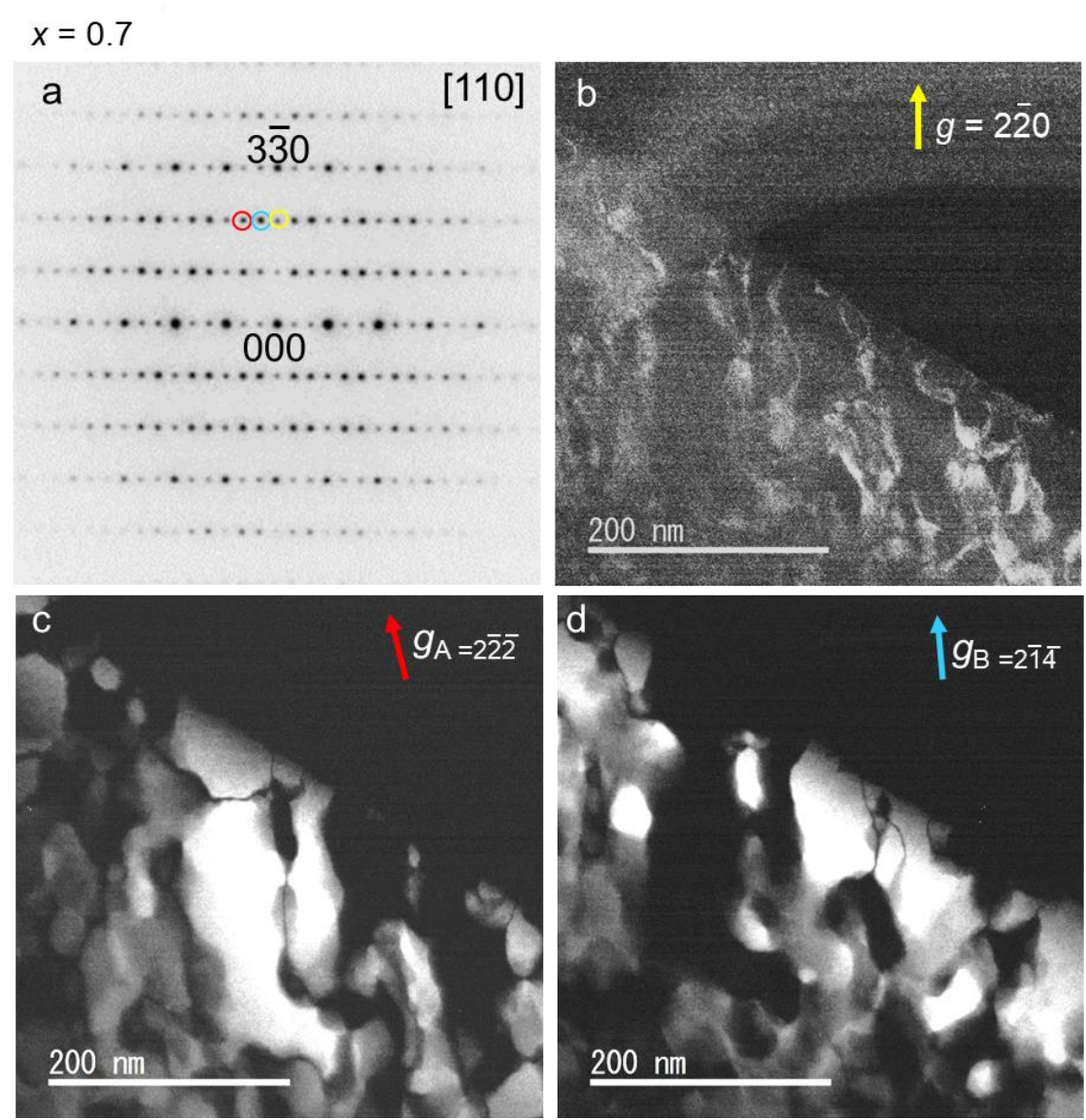

Supplementary Figure 7. Observation of microstructure in $x = 0.7$. (a) Electron diffraction pattern along [110]. The dark-field images were obtained using (b) $g = 2\bar{2}0$, (c) $g_A = 2\bar{2}\bar{2}$, and (d) $g_B = 2\bar{1}\bar{4}$, which are marked by the circles in panel (a). The reflections $g_A$ and $g_B$ originated from the twin domains A and B of Supplementary Fig. 1, respectively.

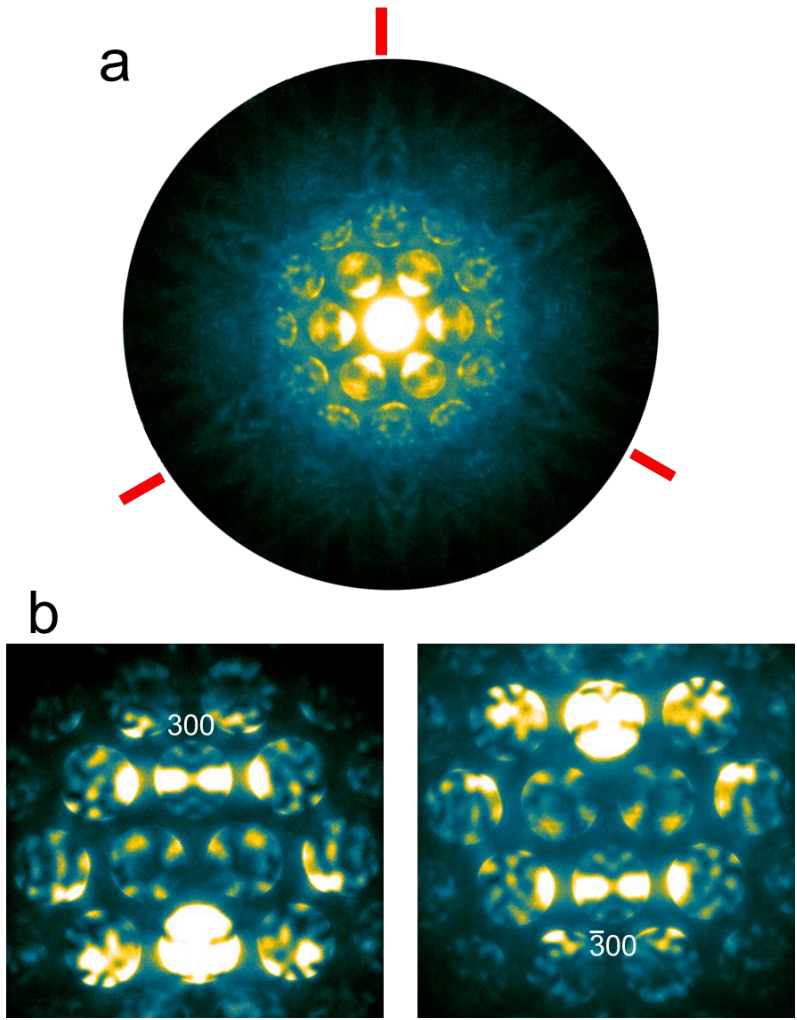

Supplementary Figure 8. Convergent-beam electron diffraction (CBED) patterns along [001] of an $R\bar{3}c$ domain in HoMn$_{1-x}$Ti$_x$O$_{3+\delta}$ ($x$ = 0.3). The probe size of the electron beam was approximately 1 nm. (a) CBED pattern of the crystal zone axis. The red bars indicate the positions of mirror symmetry. (b) ±Dark-field patterns of the 300 and $\bar{3}$00 reflection disks (±DP). The patterns show $2_R$ symmetry, which means DPs can be superimposed by parallel translation. The crystals of the point group $\bar{3}m$ and $3m$ belong to the diffraction group $6_Rmm_R$ and $3m$ in the [001] axis, respectively. Only the diffraction group $6_Rmm_R$ shows $2_R$ symmetry in ±DPs while $3m$ does not. Thus, the observation results support the space group of $R\bar{3}c$ and deny noncentrosymmetric $R3c$ in HoMn$_{1-x}$Ti$_x$O$_{3+\delta}$.

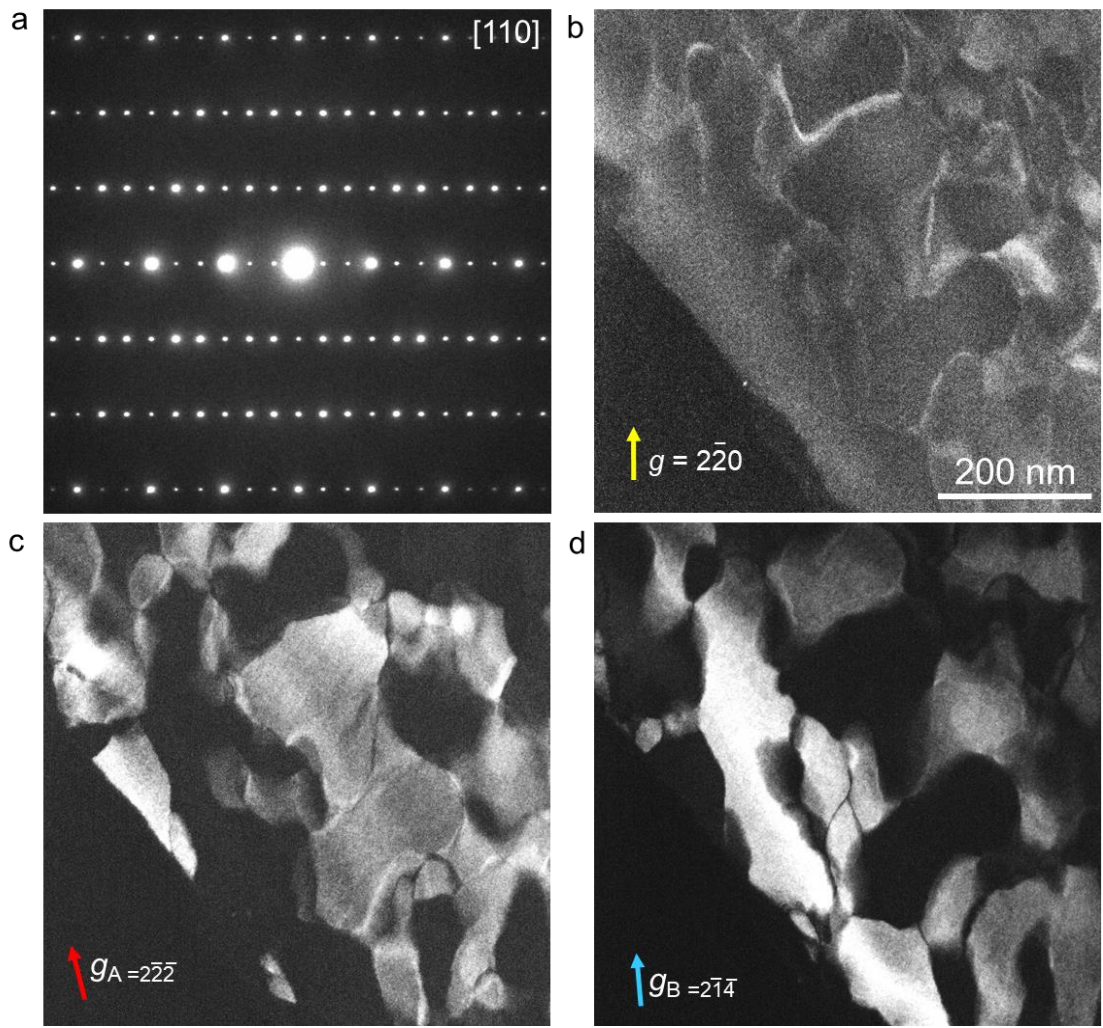

Supplementary Figure 9. Microstructure observation in the specimen ($x = 0.3$) annealed at 800 °C under the air atmosphere. In the specimen, the $R\bar{3}c$ domains first disappeared in the heating experiment. The $R\bar{3}c$ domains were formed again by the air annealing similar to the as-grown specimen. (a) Electron diffraction pattern along the [110] axis. The dark-field images using (b) $g = 2\bar{2}0$, (c) $g_A = 2\bar{2}\bar{2}$, and (d) $g_B = 2\bar{1}\bar{4}$ show that the air annealing recovers $R\bar{3}c$ twin domains and $P6_3cm$ boundaries. In (b), the bright lines correspond to $P6_3cm$ regions. The observation was performed at room temperature.